\documentclass[fleqn,12pt,twoside]{article}
\usepackage{espcrc1}
\def\be{\begin{equation}}
\def\ee{\end{equation}}
\def\bea{\begin{eqnarray}}
\def\eea{\end{eqnarray}}
\def\R{{\cal R}}
\def\J{{\cal J}}
\def\I{{\cal I}}
\def\M{{\cal M}}

\def\S{{\cal S}}
\def\vP{{\vec P}}
\def\vQ{{\vec Q}}
\def\vp{{\vec p}}
\def\vq{{\vec q}}

\def\vk{{\vec k}}
\def\vkap{{\vec \kappa}}
\def\vv{{\vec v}}
\def\vj{{\vec j}}
\def\vs{{\vec s}}
\def\<{\langle}
\def\>{\rangle}
\def\Tr{{\rm{Tr}}}
\def\half{{\textstyle{1\over 2}}}

\def\tturd{{\textstyle{2\over 3}}}
\def\fourth{{\textstyle{1\over 4}}}

\newcommand{\AmS}{{\protect\the\textfont2
  A\kern-.1667em\lower.5ex\hbox{M}\kern-.125emS}}
\title{Scaling of Hadronic Form Factors in Point Form Kinematics}

\author{F. Coester
\address{Physics Division, Argonne National
        Laboratory, \\ 
        Argonne, IL 60439, USA\\},
        D. O. Riska 
\address{Helsinki Institute of Physics and
	  Department of Physical Sciences, \\
        POB 64, 00014 University of Helsinki, Finland}}

\begin{document}

\maketitle

\begin{abstract}
The general features of baryon form factors calculated with
point form kinematics are derived.
With point form kinematics and spectator currents hadronic form factors 
are functions of
$\eta:={\textstyle{1\over 4 }}(v_{out}-v_{in})^2$ 
and, over a range of $\eta$ values, are insensitive
to unitary scale transformations of the model wave functions when
the  extent of the wave function is small compared 
to the scale defined by the
constituent mass, $\langle r^2\rangle\ll 1/m^2 $. The form factors 
are sensitive to the shape of such compact wave functions. 
Simple 3-quark proton wave functions are employed to illustrate these 
features. Rational and algebraic model wave functions
lead to a reasonable 
representation of 
the empirical form factors, while Gaussian wave functions fail.
For large
values of $\eta$ point form kinematics with spectator currents leads to
power law behavior of the form factors.

\end{abstract}

\section{Introduction}

The calculation of hadronic form factors within the framework of 
Poincar\'e covariant
quantum mechanics involves the following
two separate ingredients: (1) Bound-state wave functions, which represent
vectors in the representation space of the little group of the 
Poincar\'e group, and (2) current operators that are covariant under 
a kinematic subgroup. Covariant conserved currents are generated from 
these ingredients by the dynamics, while the choice of the kinematic 
subgroup 
specifies the form of kinematics. Spectator currents, by definition, commute with quark spectator  momenta, the definition of which depends on the choice of a form of kinematics.

With instant and front form kinematics spectator
currents provide an impulse approximation, in
which the spatial  structure of the bound-state wave function determines 
the quantitative features of the form factors. When the spatial extent 
of the bound-state wave function is scaled unitarily to zero both instant and 
front form kinematics yield form factors, which are
independent of momentum transfer.

Point-form kinematics employs the full Lorentz group 
as the kinematic subgroup. In this case the 
Lorentz covariant spectator currents probe the velocity structure
specified by the bound-state wave function. When the  extent 
of the  wave function is scaled unitarily to zero,  point-form 
kinematics yields a nontrivial scaling limit for the form factors, which 
depends 
on the shape of the wave function. At high values of momentum transfer 
the scaled form 
factors decrease with an inverse power of the momentum transfer.
The power is determined by the current operator and is independent of 
the shape of the  wave function.
A derivation of these features is presented in sections 2 and 3 below.
The purpose here is to illuminate the unfamiliar qualitative features of 
relativistic quantum mechanics with point form kinematics. Relations
to quantum field theory are beyond the scope of this article.

Recently point form kinematics \cite{Dirac,Bakam} was applied to a 
constituent quark model calculation of the form factors of the nucleon, 
which achieved
remarkable agreement with experimental data
\cite{graz1,graz2,graz3}. The calculation employed a fairly
compact 3 quark wave 
function ($\langle r^2\rangle \sim 0.1$ fm$^2$) and point-like quark 
 currents. A comparably compact 
wave function employed 
with  instant form kinematics requires either that the 
constituent quarks are spatially extended or implementation of vector
meson dominance for agreement
with the empirical form factors \cite{helminen}. 
The qualitative difference between form factors calculated
with point form kinematics and those calculated with
instant and front form kinematics 
has been recently 
emphasized in ref.\cite{desplanques}. 

The general features of  
the point form kinematics
of the confined quark description of hadron states and 
current matrix elements are described in section 2 below. 
Unitary scale transformations of the bound state
wave functions are described in section 3. Section
4 contains the illustration of the shape dependence and
the insensitivity of the proton form factors to scale changes with 
simple 3-quark wave functions for the proton. Section 5
contains a summarizing discussion. Explicit expressions
for the Wigner rotation operators, the boost  relations
of the spectator momenta to the constituent  momenta,  
and Dirac current kernels are listed in 3 appendices.
\section{Constituent quark representations of single-hadron states 
and current\\  density operators}

\subsection{Point form kinematics of confined quark.}

Single hadron eigenstates with $n$ constituents with four-momentum 
$P=M v$, may be represented by functions of the form
\be
\Psi_{M,j,v_a,\sigma}(\vv; \vk_1,\dots, 
\vk_n;\sigma_1,\dots,\sigma_n)=\phi_
\sigma(\vk_1,\dots,\vk_n;\sigma_1,\dots,\sigma_n)\delta^{(3)}(\vv-\vv_a)\; ,
\ee
where $\vk_i$ and $\sigma_i$ are constituent momenta and spin variables.
Flavor and color variables are implied.
The norm of the wave function $\phi_\sigma$ is specified by
\begin{equation}
\|\phi_\sigma\|^2=\sum_{\sigma_1,...\sigma_n}
\int d^3 k_1\ldots\int d^3 k_n \delta\left(\sum_i \vec k_i\right)
\vert  \phi_\sigma(\vec k_1,\ldots,\vec k_n;
\sigma_1,\ldots,\sigma_n)\vert^2\; ,
\label{norm}
\end{equation}
which implies that
\be
(\Psi_{M,v_f,\sigma'},\Psi_{M,v_a,\sigma})=\delta^{(3)}(\vv_f-\vv_a)
\delta_{\sigma',\sigma}.
\ee
Under Poincar\'e transformations the velocity $v$ transforms as a four-vector, 
while the total spin operator $\vj$ undergoes Wigner rotations,
$\R_W(\Lambda,v):=B^{-1}(\Lambda v)\Lambda B(v)$ as:
\be
U^\dagger(\Lambda,d)\, v\, U(\Lambda,d)= \Lambda v \; ,\qquad v^2=-1\;,\qquad
U^\dagger(\Lambda,d)\, \vj_i\, U(\Lambda,d)= \R_W(\Lambda,v) \vj_i\; .
\ee
Here the boost $B(v)$ is the rotationless Lorentz transformation,
which satisfies the
defining relation $B(v)\{1,0,0,0\}=v$.

The constituent momenta and spins undergo the same 
Wigner rotations:
\be
U^\dagger(\Lambda)\, \vk_i\, U(\Lambda)=R_W(\Lambda,v)\, 
\vk_i\; ,\qquad
U^\dagger(\Lambda)\, \vec j_i\, U(\Lambda)= \R_W(\Lambda,v)\, 
\vec j_i\; .
\label{LL}
\ee
The Poincar\'e covariance of the bound-state wave function $\phi_\sigma$,
is realized by its covariance under rotations, invariance under translations
and independence  of the velocity $v$. 
By assumption $\phi$ is an eigenfunction of an invariant mass operator. 
No constituent quark masses
are required  in this representation, but they  provide 
an essential scale in the definition of the current operators.

Some  features of the unobservable wave functions are indirectly observable
through the matrix elements 
of the covariant current
operators $I^\mu(x; v_f,v_a)$, which satisfy
the relation
\be
I^\mu(x; v_f,v_a)=e^{\imath \M\, v_f \cdot x}
\,I^\mu(0;v_f,v_a)\,e^{-\imath \M \, v_a\cdot  x},
\ee
where $\M$ is the mass operator. The mass operator of confined quark  may be 
defined by the  eigenvalues $M_n$ and assumed wave functions, $\phi_n$,
\be 
\M:= \sum_n \phi_n M_n \phi_n^\dagger\; ,
\ee
or by the conventional assumption that either $\M$ or $\M^2$ 
may be expressed as the sum of a kinetic term, 
which is a function of the internal
momenta  and a confining term, which is function of the
operators conjugate to the internal momenta. 
The basic mass operator of 
confined quark need not involve constituent  quark masses  and the formal 
structure of the dynamics is simpler if it does not \cite{feynman,dannbom}. 
This
implies that the gross features of the 
mass spectrum and the spatial extent of the wave 
function are related. Since the kinetic part of the
mass operator is repulsive it follows with this convention 
that the use of $\M$ 
 leads to more compact wave functions than the use of $\M^2$ 
\cite{helminen,dannbom,ckp,Hamil}.

The current operators 
are represented by the kernels\\ 
$
\<\sigma_1',\dots,\sigma_n ',\vk_2'\dots \vk_n'|I^\mu(0;v_f,v_a)|\vk_n,\dots
\vk_2,\sigma_n,\dots,
 \sigma_1\>\; ,
$
from which the dependent 
momentum $\vk_1:= -(\vk_2+\dots+\vk_n)$ has been omitted.
The electric and the magnetic currents,
$I^\mu_e(v_f,v_a)$ and $I^\mu_m(v_f,v_a)$,
are defined respectively by the projection 
into the plane defined  by $v_f$
and $v_a$ and  the projection perpendicular to that plane. The magnetic current
is then conserved by definition:
\be
\M\; v_f \cdot I_m(v_f,v_a) -   v_a\cdot I_m(v_f,v_a)\;\M =0,
\ee 
since $v_f \cdot I_m(v_f,v_a)=v_a\cdot I_m(v_f,v_a)=0$.

For the electric current the conservation requirement can be satisfied
by the expression
\bea
I^\mu_e(v_f,v_a)&=&\half\left( \M^\half \I_e(\eta)\M^{-\half}+
\M^{-\half}\I_e(\eta)\M^{\half}\right)
{v^\mu_f+v^\mu_a\over 2\sqrt{1+\eta}}\cr\cr &+&\half\left( \M^\half
\I_e(\eta)\M^{-\half}
-\M^{-\half}
\I_e(\eta)\M^{\half}\right){v^\mu_f-v^\mu_a\over 2}{\sqrt{1+\eta}\over \eta},
\label{ECONS}
\eea
where $\eta$ is defined as
\be
\eta:=\fourth (v_f-v_a)^2\; , \qquad -\fourth(v_f+v_a)^2= 1+\eta\; .
\ee
The Lorentz invariant operator $\I_e(\eta)$ is a functional of the current:
\be
\I_e(\eta)= {1\over 2 \sqrt{1+\eta}}
\left\{\M^{-\half} I\cdot v_f\M^{\half}+ 
\M^{\half} I\cdot v_a\M^{-\half}\right\}
\; .
\ee
The expression 
(\ref{ECONS}) may be viewed as a quantum mechanical analog of the
Ward identity.

For convenience, without loss of generality, we may assume
\be
v_a = \{\sqrt{1+\eta},0,0,-\sqrt{\eta}\}\; ,
\qquad  v_f = \{\sqrt{1+\eta},0,0,\sqrt{\eta}\}\; .
\label{BREIT}
\ee
The magnetic current then has the 
components $\{0,\I_{m\, x}(\eta),\I_{m\, y}(\eta),0\}$
and magnetic form factors are proportional to the invariant reduced matrix
elements of $\vec \I_m(\eta)$.
Electric form factors
are proportional to the invariant reduced matrix elements of $\I_e(\eta)$.
These operator relations simplify significantly when projections onto
eigenstates of $\M$ with eigenvalues $M_f$ and $M_a$ are considered.
The following treatment is restricted  to elastic transitions,
$M_f=M_a$. Electric and magnetic form factors are invariant reduced 
matrix elements of the operators  $\I_e(\eta)$ and  $\vec \I_m(\eta)$.

\subsection{Spectator currents}

Changes in the representation of initial and final states are convenient  to
accommodate the construction of simple current operators. 
Individual four-momenta $p_i$ for the spectator constituents
may be defined as functions of the $n-1$ constituent 
momenta $\vk_2,\dots,\vk_n$, a constituent
quark mass $m$ and the velocity $v$ as:
\be
p_i:= B(v)k_i\; , \qquad   k_i:=\{\omega_i,\vk_i\}\; ,\quad
\omega_i:=\sqrt{m^2+|\vk_i|^2}\; .
\label{PK}
\ee
It follows from this definition and eq.(\ref{LL}) that the momenta $p_i$ 
transform as four-vectors,
\be
U^\dagger(\Lambda)\,p_i\,U(\Lambda) = \Lambda\, p_i\; .
\ee
The parameters, which specify the boost are  different variables in 
different forms of kinematics.
With Lorentz kinematics, as used here, the components of the velocity 
$\vec v$ are the independent
kinematic variables. With other forms of kinematics the velocity, which
specifies the boost, is a
function of the kinematic components of the total momentum, the internal 
momenta $\vk_i$
and the constituent quark masses.

Free-particle spin operators $\vs_i$ which transform according to
\be
U^\dagger(\Lambda)\,\vs_iU(\Lambda)=\R_W(\Lambda, p_i)\,\vs_i\; ,
\ee
 are related to the constituent spins $\vec j_i$ by
\be
\vs_i= \R_W[B(v),k_i]\vec j_i\; .
\label{SK}
\ee
Since the relations (\ref{PK}) and (\ref{SK})  are invertible one
may choose the spectator
momenta $\vp_i$ and spin components $\lambda_i$ as the independent variables.
The transformation involves multiplication of the wave function by the square
 root of the Jacobian:
\be
\J(v,\vp_2,\dots,\vp_n) ={\partial( \vk_2,\dots,\vk_n)\over 
\partial(\vp_2,\dots,\vp_n)}
={\omega_2\cdots \omega_n\over E_2\cdots E_n}\; ,
\qquad  E_i:=\sqrt{m^2+|\vp_i|^2} \; ,
\ee
and products of  Wigner rotations $\R_W[B(v),k_i]$. In terms of these
variables the wave function takes the form
\bea
\psi(\vv;\sigma_1, \vp_2,\lambda_2,\dots,\vp_n,\lambda_n):=&&\cr\cr
\sum_{\sigma_2\dots \sigma_n}
 \prod_{i=2}^nD^\half_{\lambda_i,\sigma_i}
\Bigl(\R_W[B(v),k_i]\Bigr)                   
&\sqrt{\J(v,\vp_2,\dots,\vp_n)}&
\phi\left(\sigma_1,\vk_2[\vv,\vp_2],\sigma_2,\dots 
,\vk_n[\vv,\vp_n],\sigma_n\right),\cr &&
\eea
with the norm
\be
\Vert\psi\Vert^2 = \sum_{\sigma_1}\sum_{\lambda_2\dots\lambda_n}
\int d^3 p_2\dots \int d^3p_n
|\psi (\vv;\vp_2,\dots,\vp_n)|^2 =\Vert  \phi \Vert^2=1.
\ee

In this representation spectator currents have the general form
\bea
&&(\sigma_1,\vp_2\,',\lambda_2'\dots ,\vp_n\,',\lambda_n'|I^\mu(0;v',v)
|\vp_n,\lambda_n,\dots,\vp_2,\lambda_2,\sigma_1):= \cr\cr 
&& 
(\sigma_1'|I_1^\mu(v',v;\vp_2,\cdots,\vp_n)|\sigma_1)
\delta(\vp_2\,'-\vp_2)\cdots\delta(\vp_n\,'-\vp_n)
\delta_{\lambda_2',\lambda_2}\cdots  \delta_{\lambda_n',\lambda_n} \; .
\label{SPC}
\eea
The dependence of the current 
$I_1^\mu(\vp_2,\dots,\vp_n)=I_{1e}^\mu(\vp_2,\dots,\vp_n)
+I_{1m}^\mu(\vp_2,\dots,\vp_n)$ 
on the spectator momenta
is subject to model assumptions. The simplest form is independent of 
the spectator  momenta,
which implies structureless fermionic constituents:
\bea
(v_f, \sigma_1'|I_{1m}^\mu(\vp_2,\cdots,\vp_n)|\sigma_1,v_a)&:= &
\bar u_{\sigma_1'}(v_f)\left(\gamma^\mu -{(v_f+v_a)^\mu\over 2(1+\eta)}\right)
u_{\sigma_1}(v_a),\cr\cr
(v_f, \sigma_1'|I_{1e}^\mu(\vp_2,\cdots,\vp_n)|\sigma_1,v_a)&:= &
\bar u_{\sigma_1'}(v_f) {(v_f+v_a)^\mu\over 2(1+\eta)} u_{\sigma_1}(v_a)= 
{(v_f+v_a)^\mu\over 2 \sqrt{1+\eta}}\,\delta_{\sigma_1',\sigma_1}.
\label{spc}
\eea
In that case the charge form factor is the overlap
integral of the 
initial and final state wave functions.
The model independent $\eta$ dependence of the square of the
product of the Jacobians, which appears in the overlap
integral, is given by
\bea
\J_{fa}:=\J(v_f,\vp_2,\dots\vp_n)\J(v_a,\vp_2,\dots\vp_n)&=& 
\prod_{i=2}^n {(v_f \cdot p_i)(v_a\cdot p_i)\over E_i^2}
= \prod_{i=2}^n\left(1+ {\eta (m^2+p_{i\perp}^2)\over m^2+|\vp_i|^2}\right).
\cr &&
\label{jacob}
\eea
The Jacobian factor has an obvious zero-mass limit
\be
\left(\J_{fa}\right)_{m=0} =
\prod_{i=2}^n\left(1+\eta(1-z_i^2)\right)\; , \qquad
z_i:=p_{iz}/|\vp_i|\; ,
\label{JACZ}
\ee
which is independent
of the magnitudes of the momenta.
For large values of $\eta$ the Jacobian 
factor $\sqrt{\J_{fa}}$ is proportional
to $\eta^{(n-1)/2}$.

Since the spectator constraints (\ref{SPC}) imply the relations
\be 
k_i' = B^{-1}(v_f)B(v_a) k_i=B(v_a)^2\, k_i\; , \quad i=2,\dots ,n\; ,
\ee
between the initial and final constituent momenta, the spectator Wigner 
rotations are $\R_W[B(v_a)^2,k_i]$.  Explicit expressions 
for these operators are given in Appendix A.

For small values of $\eta$ the spectator Wigner 
rotations reduce to the identity.

Single-quark  current kernels, $\<\vp_1\,',
\lambda_1'|I^\mu(0)|\lambda_1,\vp_1\>$
and the associated Wigner rotations
introduce an additional $\eta$ dependence in
the spectator current:
\be
(v_f;\sigma_1'|I_1^\mu(\vp_2,\cdots,\vp_n)|\sigma_1,v_a):=\
\sum_{\lambda_1',\lambda_1}
D^\half_{\lambda_1',\sigma_1'}(\R_{Wf}^\dagger)
\< p_1',\lambda_1'|I_1^\mu(0)|\lambda_1,p_1\>
D^\half_{\lambda_1,\sigma_1}(\R_{Wa}),
\label{SPCUR}
\ee
with $\R_{Wa} := B^{-1}(p_1)B(v_a)B(k_1)$ 
and $\R_{Wf}:=B^{-1}(p_1')B(v_f)B(k_1')$ and 
\bea
 \< p_1'I_1^\mu(0)p_1\> =\bar u(p_1')\gamma^\mu u(p_1)\; .
\eea
The boost relations (\ref{PK}) relate  the quark momenta 
$p_1$ and $ p_1'$ to internal
momenta $\vk_1$ and $\vk_1\,'$, which are functions of the spectator momenta  
$\vk_i$ and $\vk_i\,'$. Thus the initial and final momenta $p_1$ and $ p_1'$
are boost dependent functions of $\eta$ and the spectator  momenta 
$p_2,\dots p_n$. Explicit expressions for these relations are 
given in Appendix B. 
The details of the spinor currents are in given in Appendix C.
The explicit representations of the Wigner rotations $\R_{W_a}$ and  
$\R_{Wf}$ are in Appendix A.   

\section{Unitary scale transformations}

The rms radius $r_0$ of the matter distribution represented by
the wave function $\phi$ is defined by
\begin{equation}
r_0^2 := -6 \left({dF_0(Q^2)\over dQ^2}\right)_{Q^2=0}\; ,
\label{r0}
\end{equation}
where $F_0$ is the overlap integral
\begin{equation}
F_0(Q^2):=\int d^3k_2\cdots\int d^3k_n
\phi(\vk_2-\vQ/2n,\dots,\vk_n-\vQ/2n)^*\phi(\vk_2+\vQ/2n,\dots,\vk_n+\vQ/2n).
\label{overlap}
\end{equation}
For any given wave function
the  radius $r_0$ can be varied to any positive
value by
the unitary transformation
\be
U_\beta \phi(\vk_2,\dots\vk_n)
= \beta^{-3(n-1)/2}\phi(\vk_2/\beta,\dots,\vk_n/\beta)\; .
\ee
The radius $r_0$ is a measure of the extent of the Fourier transform of the 
wave function. In the limit $\beta \to \infty$ the radius $r_0$ is reduced to
zero. Quark masses appear as scale parameters in the currents. 
With zero mass constituents
the current  operators  commute with the unitary scale
transformations $U_\beta$.
With point-form kinematics the
relevant structure is the distribution of internal velocities $\vk_i/m$.
The dimensionless form factors are functions of $\eta$ and $mr_0$,
and in the zero mass limit are invariant under unitary scale
transformations. The point limit $r_0 \to 0\,$
and the zero-mass limit are identical. However, when the mass operator  $\M$
(or $\M^2$) of confined quark is the sum of a kinetic term plus a 
confining potential the scale $r_0$ is related to the mass spectrum 
independently of a constituent mass.
 
With ``Galilean relativity'' $r_0$ is equal to the 
observable charge radius. With instant and front-form
kinematics, and $m\,r_0>1$, the 
relation between the radius  
$r_0$  and the charge operator involves
``relativistic corrections'' , which arise from the
boosts and the spinor structure.
At this point the qualitative difference between point
and instant-form  kinematics is readily apparent. 
In the latter form the  kinematic quantity that specifies  the total
momentum is $\vP$ instead of $\vec v$. Constituent momenta $p_i$ are
specified  as functions of $\vP$ and $\vk_2,\dots,\vk_n$ by
\bea 
p_i&:=&- B(-\vQ/2M_0)\{\omega_i,\vk_i\}\; ,
\qquad M_0:=\sum_{i=1}^n\omega_i \; , \cr\cr
p_i'&:=&B(\vQ/2M_0')\{\omega_i',\vk_i\,'\} 
\qquad M_0':=\sum_{i=1}^n\omega_i'\; . 
\eea
It follows from this definition that
\be
\sum_i \vp_i =-\half \vQ\;, \qquad \mbox{and}\qquad \sum_i \vp_i\,' 
=\half \vQ\;
\ee
The spectator constraints are $\vp_i\,'=\vp_i$ for $i=2,\dots,n$.
Since 
\be
\lim_{\beta\to \infty}U^{-1}(\beta){\vQ\over M_0}U(\beta) =0,
\ee
the spectator constraints reduce to $\vk_i\,'-\vk_i$ and the form factors are
independent of $\vQ$.

With Lorentz kinematics  
the charge form factors decrease with a  power  
of $\eta$ in the scaling limit, $ \beta \to \infty$,
when $\eta\gg 1$ . The exponent  is independent of the shape of the wave function. 
This  becomes evident with the  change
of variables of integration variable $\vp_i \to \sqrt{\eta}\, \vp_i$,
which  brings a factor of
$\eta^{-3(n-1)/2}$ in front of the integral. Combined with the
asymptotic $\eta$ dependence of the Jacobian factor (\ref{jacob})
the overlap integral is proportional to $\eta^{-(n-1)}$ 
at large values of $\eta$. Dirac spinor currents (\ref{SPCUR})
introduce an additional asymptotic $\eta$ dependence of the form factor, 
proportional to $1/\eta^{3/2}$.
Consequently point form kinematics with the single particle current
kernel (\ref{SPCUR}) implies that the electric
form factors behave as $\eta^{-(n+1/2)}$ at large values of $\eta$.
\section{Proton form factor illustrations}

\subsection{Expressions for $G_E$ and $G_M$}

In order to illustrate the  behavior
of the form factors we employ a conventional
completely symmetric spin-isospin amplitude
$\chi_{\sigma,\tau}$ represented by a 
function of Clebsch-Gordan coefficients:
\bea
\chi_{\sigma,\tau}(\sigma_1,\tau_1,\sigma_2,\tau_2,\sigma_3,\tau_3)&:=&
{1\over \sqrt{2}}\Bigl\{ \delta_{\sigma,\sigma_1}
(\half,\half\sigma_2,\sigma_3|0,0)(\half,\half\tau_2,\tau_3|0,0)
\cr\cr
&+& (\half,\half\sigma_2,\sigma_3|1,\sigma_2+\sigma_3)
(1,\half,\sigma_2+\sigma_3,\sigma_1|\half, \sigma)\cr\cr
&\times&
(\half,\half\tau_2,\tau_3|1,\tau_2+\tau_3)
(1,\half,\tau_2+\tau_3,\tau_1|\half, \tau)\Bigr\}\; ,
\eea
and 
two different permutation symmetric
radial  wave function models. These have the Gaussian 
and rational shapes:
\be
\phi_G\left({\kappa^2+q^2)\over 2 b^2}\right):={1\over (b\sqrt{\pi})^3}
\exp\left(-{\kappa^2+q^2\over 2 b^2}\right),
\label{Gaussian}
\ee
\be
\phi_R\left({\kappa^2+q^2\over 2 b^2}\right):=b^{-3}\sqrt{3\over
4\pi^3}\left(1+{\kappa^2+q^2
\over 2 b^2}\right)^{-2},
\label{rational}
\ee
where
\be
\vkap:=\sqrt{{3\over 2}} 
\vk_1 \equiv -\sqrt{{3\over 2}}(\vk_2+\vk_3)\; ,\qquad
\vq:= \sqrt{\half}(\vk_2-\vk_3)\; ,\qquad{\partial(\vkap,\vq)\over
\partial(\vk_2,\vk_3)}=\sqrt{27}\; ,
\label{KAPQ}
\ee
and
\be
\half(\vkap^2 +\vq^2)= \vk_2^2+\vk_3^2+\vk_2\cdot\vk_3
\equiv \half(\vk_1^2+\vk_2^2+\vk_3^2)
\; .
\ee
Changes of the scale parameter $b$ represent unitary transformations.
The overlap integral $F_0(Q^2)$ (\ref{overlap}) takes the convenient form
\be
F_0(Q^2)=\int d^3\kappa'\int d^3\kappa \int d^3 q 
\phi\left({{\kappa'}^2+q^2\over 2 b^2}\right) 
\phi\left({\kappa^2+q^2\over 2 b^2}\right)
\delta^{(3)}\left(\vkap'-\vkap-\sqrt{\tturd}\,\vQ\right)).
\ee

By definition the proton  form factors $G_E(\eta)$ and $G_M(\eta)$ are related
to the electric and magnetic current matrices  by
\bea
G_E(\eta)&=& \half \Tr(\psi_f \I_e(\eta)\psi_a)\; , \cr\cr
G_M(\eta)&=&\half  \Tr[(\sigma_x-\imath\sigma_y)(\psi_f,\I_{m+}\psi_a)].\qquad \I_{m+}:=\I_{mx}+\imath \I_{my}.
\eea
After summation over the spin-isospin indices
the expressions for the   factors of the proton reduce to the integrals
\bea
G_E(\eta)&=&\int d^3 p_2 d^3 p_3 
\phi\left({{\kappa'}^2+q^2\over 2 b^2}\right)
\phi\left({\kappa^2+q^2\over 2 b^2}\right)
\sqrt{27\J_{fa}(\vp_2,\vp_3)} \cr\cr
&&C_{23}(\eta,\vp_2,\vp_3)\S_e(\eta,\vp_2,\vp_3),
\cr\cr\cr
G_M(\eta)&=&\int d^3 p_2 d^3 p_3 
\phi\left({{\kappa'}^2+q^2\over 2 b^2}\right)
\phi\left({\kappa^2+q^2\over 2 b^2}\right)
\sqrt{27\J_{fa}(\vp_2,\vp_3)} \cr\cr
&&C_{23}(\eta,\vp_2,\vp_3)\S_m(\eta,\vp_2,\vp_3)\; .
\label{GEM}
\eea
For zero-mass constituents and  large $\eta$ the 
Jacobian factor (\ref{JACZ}) is proportional to
$\eta^2$:
\be
\J_{fa}(\eta, \vp_2,\vp_3)\approx \eta^2(1-z_2^2)(1-z_3^2).
\ee
The coefficient $C_{23}(\eta,\vp_2,\vp_3)$ is 
determined by the spectator Wigner
rotations,
\bea
C_{23}(\eta,\vp_2,\vp_3)&=&\half \sum_{\sigma',\sigma}
D^\half_{\sigma',\sigma}\left(\R_W[B(v_a)^2, k_2]\right)
D^\half_{-\sigma',-\sigma}\left(\R_W[B(v_a)^2, k_3]\right)\cr\cr &=&
\cos{\theta_2\over 2}\,\cos{\theta_3\over 2} 
+ {\vp_{2\perp}\cdot \vp_{3\perp}\over |\vp_{2\perp}||\vp_{3\perp}|}\,
\sin{\theta_2\over 2}\,\sin {\theta_3\over 2}\; .
\eea
For zero-mass constituents and large $\eta$ this coefficient 
is independent of $\eta$ (\ref{THZ23}):
\be
C_{23}(\eta,\vp_2,\vp_3)\approx {\vp_{2\perp}
\cdot \vp_{3\perp}\over |\vp_{2\perp}||\vp_{3\perp}|}\; .
\ee
For zero-mass constituents and $\eta\gg 1$ the arguments of the wave function
have the approximate forms:
\be
|\vk_i|^2 \approx \eta |\vp_i|^2(1+z_i)^2 \; ,
\qquad  |\vk'_i|^2 \approx \eta |\vp_i|^2(1-z_i)^2 
\qquad  i= 2,3,
\ee
and
\be
\vk_2\cdot\vk_3 \approx \eta |\vp_2|\vp_3| \sqrt{(1+z_2)(1+z_3)}\; , \qquad
\vk'_2\cdot\vk'_3 \approx \eta |\vp_2|\vp_3| \sqrt{(1-z_2)(1-z_3)}\;.
\ee
When the current is 
specified by (\ref{edirac}) and (\ref{mdirac}) the current factors
$\S_e $ and $\S_m$  are  
\bea
&\S_e&=
\sqrt{{(E_1'+m)(E_1+m)(1+\eta)\over 4 E_1' E_1}}
\bigg\{\Bigl(1+ {\vec p_1\,'\cdot \vec p_1 \over 
(E_1'+m)(E_1+m)} \Bigr) \cos\left({\theta_1-\theta_1'\over 2}\right)
\cr\cr && 
+{|p_{1\perp}|(p_{1z}' -p_{1z}) \over
 (E_1'+m)(E_1+m)}\sin\left({\theta_1-\theta_1'\over 2}\right)  \bigg \}\;,
\label{Se}
\eea
and
\bea
&\S_m&= 
 \sqrt{{1+\eta \over  4\eta E_1' (E_1'+m) E_1 (E_1+m)}}
\bigg\{
\Bigl[ p_{1z}'(E_1+m )- p_{1z} (E'_1+m) \Bigr]
\cos{\theta_1'\over2}\cos{\theta_1\over 2}\cr\cr
&+&|p_{1\perp}|(E_1'+E_1+2m)\sin{\theta_1'-\theta_1\over2}
+|p_{1\perp}|(E_1-E_1')\sin{\theta_1'+\theta_1\over2}\bigg\}.  
\label{Sm}
\eea
For zero-mass constituents these expressions reduce to
\bea
\S_e(m=0)&=& \sqrt{1+\eta\over 4}\biggl
\{ (1+\hat p_1'\cdot \hat p_1)
\cos\left({\theta_1-\theta_1'\over 2}\right)\cr\cr
&+&\left(\sqrt{1-z_1'^2}z_1'-\sqrt{1-z_1^2}z_1\right)
\sin\left({\theta_1-\theta_1'\over
2}\right)\biggr\},\cr\cr 
\S_m(m=0)&=&\sqrt{1+\eta\over 4 \eta}
\bigg\{(z_1'-z_1)\cos{\theta_1'\over2}\cos{\theta_1\over 2}\cr\cr
&+& \left(\sqrt{1-z_1^2}+\sqrt{1-{z_1'}^2}\right)
\sin{\theta_1'-\theta_1\over2}
+\left(\sqrt{1-z_1^2}-\sqrt{1-{z_1'}^2}\right)\sin{\theta_1'+\theta_1\over2}
\bigg\}.
\cr\cr
&&
\eea

For $\eta\gg  1$  the factor ${\cal S}_e$ is proportional to 
$\eta^{-3/2}$ and
the factor ${\cal S}_m$ is independent of $\eta$.
It then follows from 
eq. (\ref{GEM}) 
that electric and the magnetic
form factor behave as 
$\eta^{-7/2}$ and      
$\eta^{-2}$ respectively for large values of $\eta$.
With the current (\ref{spc}) one has $\S_e=\S_m=1$. In this
case both  form factors  behave as $\eta^{-2}$. From the
Rosenbluth formula  for the elastic cross section,
\be
{1\over \sigma_{Mott}}{d\sigma\over d\Omega}
={1\over 1+\eta} G_{ep}^2 +\left\{{1\over 1+\eta}
+2 \tan^2 \theta_e/2\right\}\eta G_{mp}^2,
\ee
it follows that 
the magnetic form factor dominates for large $\eta$.

\subsection{Numerical results}

For the two shapes of the wave function considered here
the rms radius (\ref{r0}) is  $r_0=1/b$ for the
Gaussian model  (\ref{Gaussian}), and $r_0 =\sqrt{2/5}/b$ for
the  rational wave function (\ref{rational}).
With $b$ = 650 MeV the Gaussian model (\ref{Gaussian}) 
gives
same mean square radius ($\sim$ 0.1 fm$^2$) as the wave function 
derived in ref.~\cite{Hamil} by diagonalization of a 3 quark mass operator, 
which gives a satisfactory description of the empirical 
nucleon spectrum. This wave function was used in refs.~\cite{graz1,graz2,graz3}
 to calculate the nucleon form factors
with point form kinematics.  
In the case of the rational wave 
function (\ref{rational}) the same value obtains with $b=$ 410 MeV.
With  the quark mass $m=$ 340 MeV used in refs.\cite{graz1,graz2,graz3} 
the relevant dimensionless parameter is $mr_0=.52$.

To illustrate the dependence of the form factor on unitary scale 
transformations of the wave function we show numerical results for
$mr_0\,=\, .52,\, .33 ,\, 0$ and both wave function shapes in Fig.\,1.
The results reveal the relative insensitivity
of the form factors to unitary scale transformations of the 
wave function  when $(mr_0)^2\ll 1$. A  recent parameterization of form
factor  data  ref.\cite{lomon} provides a bench mark 
for comparison.

With the spinor currents the magnetic form 
factors, Fig.\, 2, show a similar more compact pattern.
In that case the Gaussian wave  function in the 
point limit and  the rational wave function with
$mr_0=.52$  are  in rough agreement with each other and the data 
parameterization.
The rational wave function with $mr_0=.52$ gives the magnetic moment as
2.86 nm, which is close to the empirical value (2.79 nm). 

The rational wave function with $mr_0=.52$ provides a 
reasonable representation 
of the data, which is comparable to the results obtained with the
wave function employed in refs.\cite{graz1,graz2}.
Even in the point limit the Gaussian shape does not yield both
form factors close to the data.

With $\S_e=1$ a regular scaling  pattern obtains as shown in Figs.\,3
and 4,
which depends only on the shape of the wave function and the 
point-form spectator 
constraint.  For $m r_0 <.5$ the two
shapes  provide form factors  that range over limited  
non-overlapping regions of size.
These  figures illustrate  the effects of the Lorentz-kinematic 
spectator constraints without
the single-quark spinor currents.
With the spinor current the electric form factors 
converge more rapidly to the point limit and show
a more drastic dependence on the shape of the wave function.

\section{Discussion}

The main results of the present investigation can be summarized as 
follows:
With zero-mass constituent quarks relativistic quantum mechanics with 
point-form kinematics
yields hadron form factors, which are invariant under unitary scale 
transformations, but which
depend strongly on the shape of the wave function. 
With non-vanishing 
constituent quark masses  the form factors depend but weakly on 
the scale 
when $\langle r^2\rangle m^2$  is about
one fourth or less. Within that range it is possible to achieve
realistic features 
of the proton form factors
with simple rational wave function shapes. 
It has also been shown is that a realistic mass spectrum is compatible with 
these features \cite{graz1,graz2}. With the simple dynamical structure 
discussed in ref.\,\cite{dannbom} adjustment of the confinement shape and the quark masses 
may provide a realistic description of both the spectrum and the form factors.

The quark momenta or velocities used in the definition of currents
are related to the internal relative momenta by boost relations,
which depend the form of kinematics. The special features of Lorentz
kinematics are  numerically
significant, and lead  to the non-trivial point limits of the
form factors. The numerical significance
of the Wigner rotations of the spins  is small compared to that
of the boost dependence of the momenta.
When  
$\eta \gg 1$ the form factors in the point limit 
asymptotically attain
power law behavior. The exponent of the leading power depends on the current
model, but not on the wave function.

The results shown in Fig.\,1 suggest that it should be possible
to approximate electric form factor data with
a simple wave function  in the point limit.
Indeed we find that with
the spinor current considered above the
wave function 
\be
\phi_I(\kappa^2+q^2)=C_a\bigg(
1+{\kappa^2+q^2\over 2 b^2}\bigg)^{-a},
\label{irrat}
\ee
where $C_a$ is a normalization constant, and $a=11/4$ yields the electric form 
factor shown in Fig. 5., which is quite close to the
parameterization of ref.\cite{lomon}.
The magnetic form calculated with this wave function
in the point limit is however not as close to the
corresponding parameterization given in ref.\,\cite{lomon}.
The results shown in Fig 1 suggest that better
representations of the empirical magnetic form factor
call for moderate finite values 
of $m/b$. As examples the results for both $G_E$ and
$G_M$ as obtained with $m/b=0.52$ with the wave function
(\ref{irrat}) with $a=9/4$ are also shown in Fig. 5.
In this case the calculated electric form factor is effectively
indistinguishable from the corresponding parameterization
of the empirical values, and the calculated magnetic moment 
2.80 nm coincides with the empirical value. The calculated
magnetic form factor falls slightly faster than the
parameterization of the empirical form factor. 

The main qualitative features of relativistic quantum
mechanics with Lorentz kinematics have been illustrated
above. These features appear to be appropriate
for a phenomenology of confined quark. It appears that
the point limit, which is characteristic of Lorentz kinematics 
may provide a useful zero-order description to be refined by 
finite quark masses, which enter as scale parameters.

\section{Acknowledgment}

We are grateful for instructive discussions and
correspondence with L.Ya. Glozman,  
W. H. Klink and W. Plessas on the subject matter.
D .O. R. is indebted to R. D. McKeown for hospitality
at the W. K. Kellogg Radiation Laboratory.
Research supported in part by the Academy of Finland through
grant 54038 and by the U.S. Department of
Energy, Nuclear Physics Division, contract
W-31-109-ENG-38.

\appendix
\begin{center}
{\bf Appendices}
\end{center}

\section {Wigner rotations}

  The explicit representations of the spectator Wigner rotations 
$\R_W[B(v_a)^2,k_i]$.
are 
\be
D^\half \Bigl(\R_W[B(v_a)^2,k_i] \Bigr)=\cos{\theta_i\over 2}-
\imath \sin{\theta_i\over 2}\,
{(\vp_i\times \vec \sigma_i)_z\over |p_{i\perp}|} \; , \qquad i= 2,\dots,n .
\label{RWSP}
\ee
where the angles $\theta_i$ are defined as 
\be
\sin{\theta_i\over 2}:={-\sqrt{\eta}|p_{i\perp}|\over
\sqrt{2m(m+\sqrt{1+\eta}E_i)+|\vp_i|^2+\eta(m^2+p_{i\perp}^2)}}.
\ee
For zero-mass constituents (m=0) the Wigner rotations are independent
of the magnitudes of the momenta,
\be
\left(\sin^2 {\theta_i\over 2}\right)_{m=0}= 
{\eta(1-z_i^2)\over 1+\eta(1-z_i^2)}= 1- {1\over1+\eta(1-z_i^2)} \;,
\label{THZ23}
\ee
with $z_i:=p_{iz}/|\vp_i|$. 
The explicit representations of the Wigner rotations $\R_{W_a}$ and  
$\R_{Wf}$  associated with the single-quark spinor current are
\bea
D^\half\Bigl(\R_W[B(v_a),\vk_1]\Bigr)&=
&\cos {\theta_1\over 2}-\imath \sin {\theta_1\over2}
{ (p_1\times \vec \sigma_1)_z\over |p_{1\perp}|},\cr\cr
D^\half\Bigl(\R_W[B(v_f),\vk_1\,']\Bigr)&=
&\cos {\theta_1'\over 2}-\imath \sin {\theta_1'\over2}
{ (p_1\times \vec \sigma_1)_z\over |p_{1\perp}|}.
\eea
Here the angles $\theta_1, \theta_1 '$ are defined as
\bea
\sin{\theta_1\over2}&=&{\sqrt{\eta} |p_{1\perp}|
\over\sqrt{ 2(1+\sqrt{1+\eta})(m+E_1)(m+\omega_1)}},\cr\cr 
\sin{\theta_1'\over2}&=&-{\sqrt{\eta} |p_{1\perp}|
\over\sqrt{ 2(1+\sqrt{1+\eta})(m+E'_1)(m+\omega'_1)}}\; .
\eea
For $m=0$ and  $\eta\gg 1$ these expressions reduce to 
\be
\left(\sin{\theta_1\over 2}\right)^2_{m=0} 
\approx  1+{k_{1z}\over |\vk_1|}\; ,
\qquad\left(\sin{\theta_1'\over 2}\right)^2_{m=0}  \approx
1-{k'_{1z}\over |\vk_1'|}.
\label{THZ1}
\ee

\section {Boost relations}
According to the definition (\ref{PK})  the Breit-frame (\ref{BREIT})  
components of the spectator momenta $p_i$ are related to the initial and final 
constituent momenta $\vk_i$ and $\vk_i' $ by
\bea
\{E_i,\vp_i\}&=&\{ \sqrt{1+\eta}\,\omega_i-\sqrt{\eta}\, 
k_{iz}\,,\,k_{i\perp}\,,\,\sqrt{1+\eta}\,k_{iz}-\sqrt{\eta}\,\omega_i\} 
\cr\cr
&=&\{ \sqrt{1+\eta}\,\omega_i'+\sqrt{\eta}\,k_{iz}'\,,\,k_{i\perp}'\, ,\,
\sqrt{1+\eta}\,k_{iz}'+\sqrt{\eta}\,\omega_i'\}.
\label{SPP}
\eea
These relations can be inverted to give
\begin{eqnarray}
&&k_{i\perp}=p_{i\perp},
\quad k_{iz}=\sqrt{1+\eta}\,p_{iz}+\sqrt{\eta} E_i\; , \qquad
\omega_i=\sqrt{\eta} p_{iz}+\sqrt{1+\eta}\,E_i ,\cr\cr
&&k_{i\perp}'=p_{i\perp},
\quad k_{iz}'=\sqrt{1+\eta}\,p_{iz}-\sqrt{\eta} E_i\;
,\qquad  
\omega_i'=-\sqrt{\eta} p'_{iz}+\sqrt{1+\eta}E'_i.
\label{ks}
\end{eqnarray}
The momenta $p_1$ and $p'_1$ of the active quark are functions of  $\eta$ and the spectator momenta.
The formal relation  to the internal momenta
$\vk_1$ and $\vk_1'$ are the same as those given  in eq.  (\ref{SPP}),
\bea
p_1&:= 
&B(v_a)\{\omega_1,\vk_1\}= \left\{ \sqrt{1+\eta}\,\omega_1-\sqrt{\eta}\, 
k_{1z}\, ,\,\;k_{1\perp}\,,\,
\sqrt{1+\eta}\,k_{1z}-\sqrt{\eta}\,\omega_1\right\}\; , 
\cr\cr 
p'_1&:= &B(v_f)\{\omega_1',\vk_1' \}=
\left\{ \sqrt{1+\eta}\,\omega_1'+\sqrt{\eta}\, 
k'_{1z}\, ,\,\;k_{1\perp}\,,\,\sqrt{1+\eta}\,k'_{1z}+\sqrt{\eta}\,
\omega'_1\right\} \; ,  
 \eea
where 
\be
\vk_1:= - \sum_{i=2}^n\vk_i\; 
, \qquad \mbox{and}\qquad\vk'_1:=- \sum_{i=2}^n\vk_i'. 
\ee
It follows that 
\be
{p_{1z}\over E_1}= {\sqrt{1+\eta}\,k_{1z}-\sqrt{\eta}\,\omega_1
 \over \sqrt{1+\eta}\,\omega_1-\sqrt{\eta}\, k_{1z}}
\approx -1 +{1\over 2\eta}\,{\omega_1+k_{1z}\over \omega_1-k_{1z}},\qquad
\mbox{for}\quad \eta\gg 1,
\ee
and 
\be
{p'_{1z}\over E'_1}= {\sqrt{1+\eta}\,k'_{1z}+\sqrt{\eta}\, \omega'_1
 \over \sqrt{1+\eta}\,\omega'_1+\sqrt{\eta}\, k'_{1z}}
\approx 1 -{1\over 2\eta}\,{\omega'_1-k'_{1z}\over \omega'_1+k'_{1z}} \; ,
\qquad \mbox{for}\quad \eta\gg 1\;.
\ee
The  momenta $\vk_1$ and $\vk'_1$ are related  to
the spectator momentum
\be
p_{sp}:=\{E_{sp},\vp_{sp}\}= \sum_{i=2}^n\{E_i,\vp_i\},
\ee
by the boost relations
\be
-k_{1z}=\sqrt{1+\eta}\, p_{spz}+\sqrt{\eta}\,E_{sp}\; ,\qquad 
-k'_{1z}=\sqrt{1+\eta}\, 
p_{spz}-\sqrt{\eta}\, E_{sp} \; .
\ee

\section{Spinor currents}
For the  single-quark  Dirac current we have 
\bea
&&{\<\vp_1\,'|\I_{e1}|\vp_1\>\over \sqrt{1+\eta}}=
 {1+\beta_1\over 2}{(\vec \alpha_1 \cdot \vp_1\,'+E_1'+m)(\vec
\alpha_1\cdot \vp_1+E_1+m)\over 
\sqrt{4 E_1'(E_1'+m)E_1(E_1+m)}}{1+\beta_1\over 2}\cr\cr\cr
&=&\sqrt{{(E_1'+m)(E_1+m)\over 4 E_1' E_1}}
\left(1+{\vec p_1\,'\cdot \vec p_1\over (E_1'+m)(E_1+m)}\right)
+{\imath (\vp_1\,'-\vp_1)\cdot(\vp_1 \times\vec \sigma_1)\over
 \sqrt{4E_1'(E_1'+m)E_1(E_1+m)}}\; ,
\label{edirac}
\eea
and

\bea
&&{\<\vp_1\,'| \I_{1m+}|\vp_1\>\over \sqrt{1+\eta}}=
 {1+\beta_1\over 2}{(\vec \alpha_1 \cdot \vp_1\,'+E_1'+m)
\alpha_{1+}
(\alpha_1\cdot \vp_1+E_1+m)\over 
\sqrt{4 E_1'(E_1'+m)E_1(E_1+m)}}{1+\beta_1\over 2}
\cr\cr\cr
&=& { \sigma_{1+} \left[p_{1z}'(E_1+m)-p_{1z} (E_1'+m)\right]
\over 2\sqrt{E_1'(E_1'+m)E_1(E_1+m)}}
- {  \sigma_{1z}p_{1+} (E_1- E_1')
\over  2\sqrt {E_1'(E_1'+m)E_1(E_1+m)}}\cr\cr
&+& {p_{1+} (E_1+E_1' +2m)
\over  2\sqrt {E_1'(E_1'+m)E_1(E_1+m)}}.
\label{mdirac}
\eea
In the zero-mass limit the current kernels reduce to
\bea
\left({\<\vp_1\,'|\I_{e1}|\vp_1\>\over 
\sqrt{1+\eta}}\right)_{m=0}&=&\half \left\{ (1+\hat p_1'\cdot\hat p_1)
+\imath (\hat p_1'\times\hat p_1)\cdot \vec \sigma_1\right\} \; ,
\qquad\hat p_1:= \vp_1/|\vp_1|,
\cr\cr\cr
\left({\<\vp_1\,'|
\I_{1+}|\vp_1\>\over \sqrt{1+\eta}}\right)_{m=0}&=& 
 -{\imath\over2}\left\{  \sigma_{1+}  ( \hat p_{1z}'-\hat p_{1z})
+ p_{1+}\sigma_{1z}{|\vp_1\,'|-|\vp_1|\over |\vp_1\,'|\,\,|\vp_1|}
\right\} \; .
\eea

\centerline{Figure Captions}

\bigskip

Fig.1 The proton electric form factor calculated with point
form kinematics with the Gaussian (``G'') (\ref{Gaussian})
and rational (``R'')(\ref{rational}) wave function models
for different values of $mr_0$. The value $mr_0=0$ is
the point limit. The  parameterization (``LOMON'') 
of the data 
is taken from ref.\cite{lomon}. 

\bigskip

Fig.2  Illustration of the dependence of the proton magnetic 
form factors, 
with the spinor spectator current,
on the shape and scale of the wave function. The notation
is the same as in Fig.1. 
The  parameterization (``LOMON'') of the data 
is taken from ref.\cite{lomon}. 

Fig.3 Comparison of form electric proton form factors as
calculated with the rational wave function  (\ref{rational})
with and without (``$S_e=1$'') the spinor
factor in the current operator. The 
parameterization of the data  
``LOMON'' is that given in ref.\,\cite{lomon}.

\bigskip

Fig.4  Comparison of  electric  proton form  factors as
calculated with the Gaussian wave function (\ref{Gaussian})
with and without (``$S_e=1$'') the spinor
factor in the current operator. The 
parameterization ``LOMON'' 
of the data is that given in ref.\,\cite{lomon}.

\bigskip

Fig.5  The electric and magnetic form factors given by the 
wave function model (\ref{irrat}) with $a=11/4$ in the
point limit and with $a=9/4$ with $m/b=0.52$.
The parameterization (``LOMON'') of the data is
from ref.\cite{lomon}.

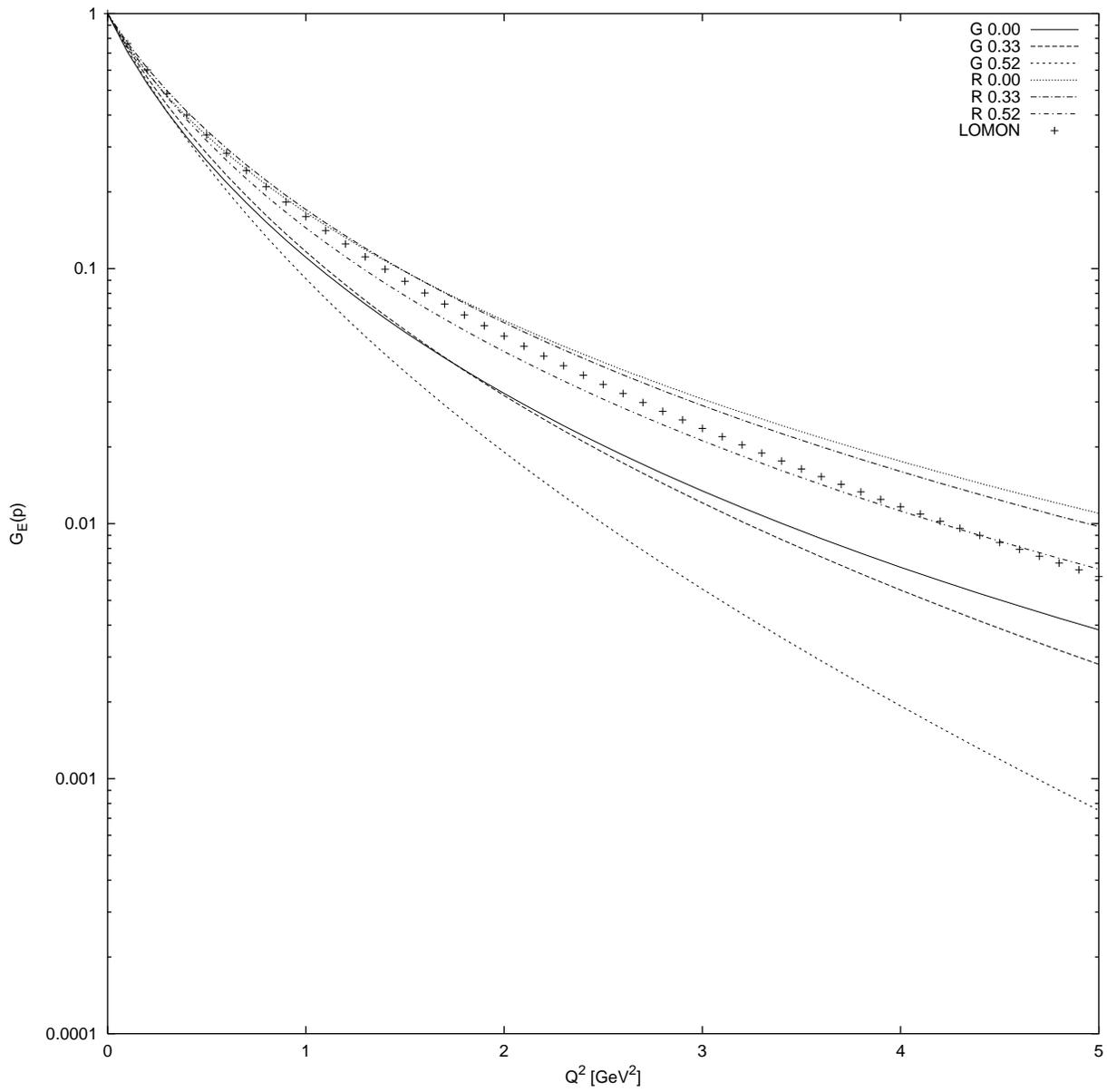
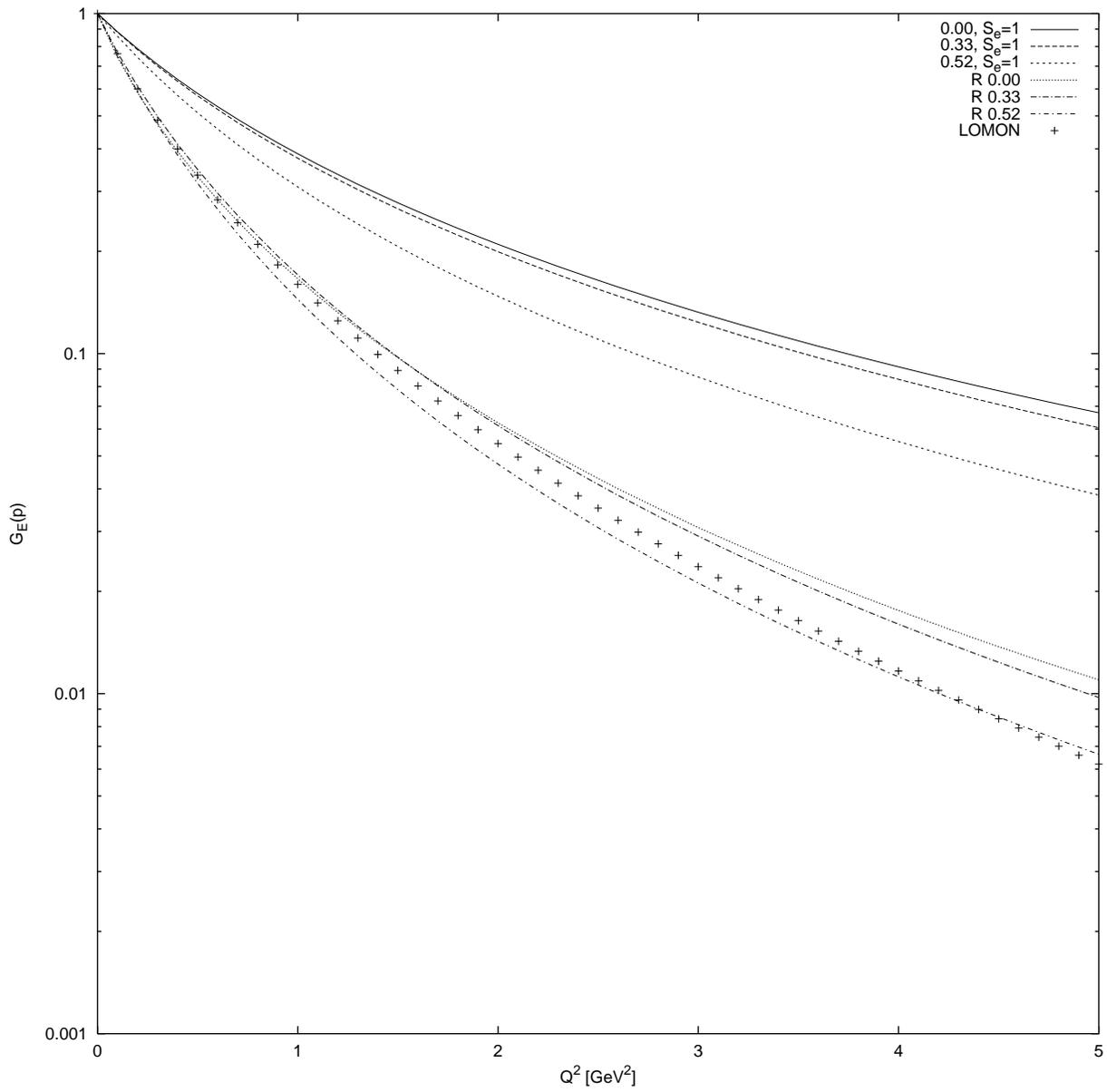
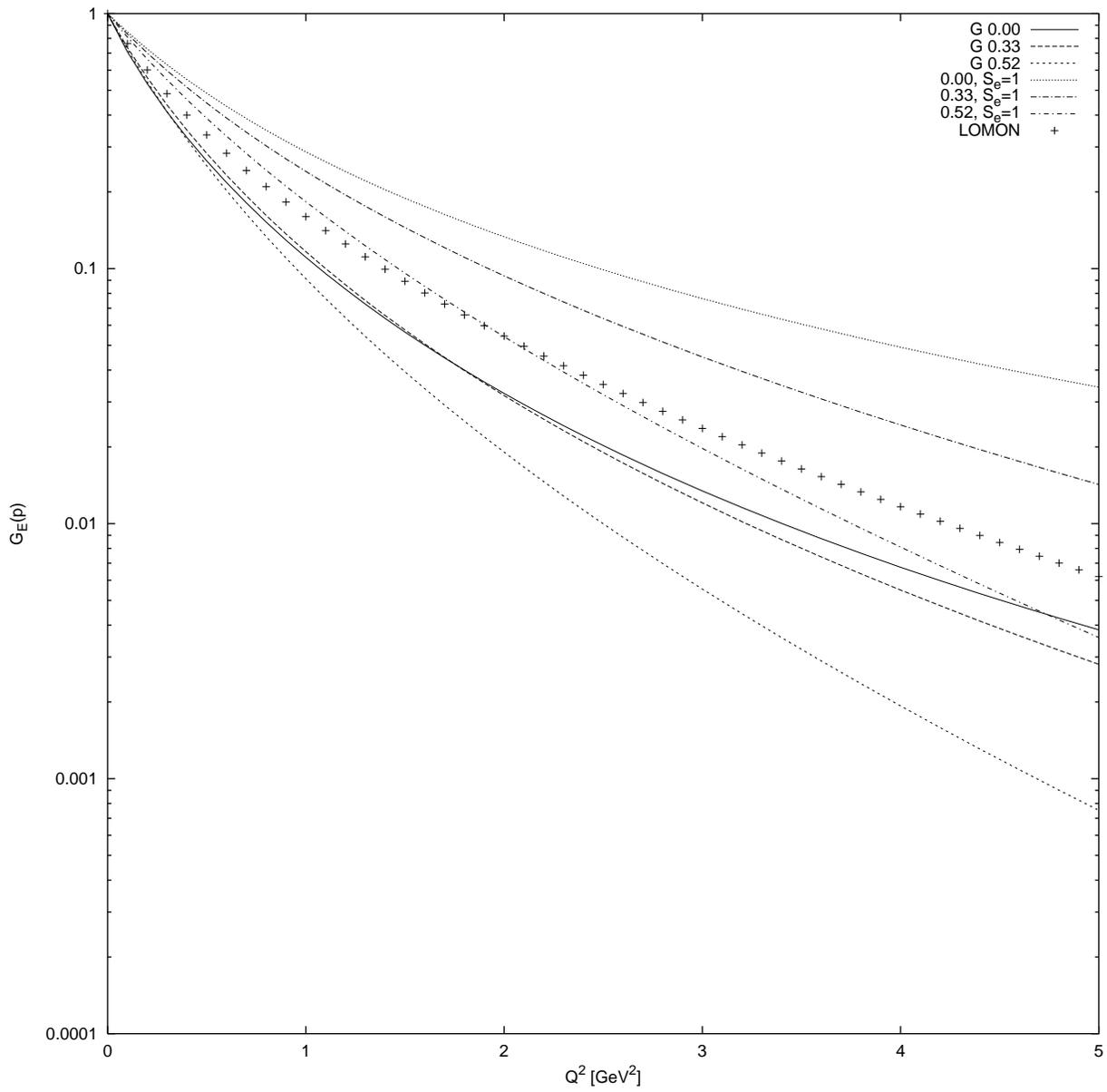
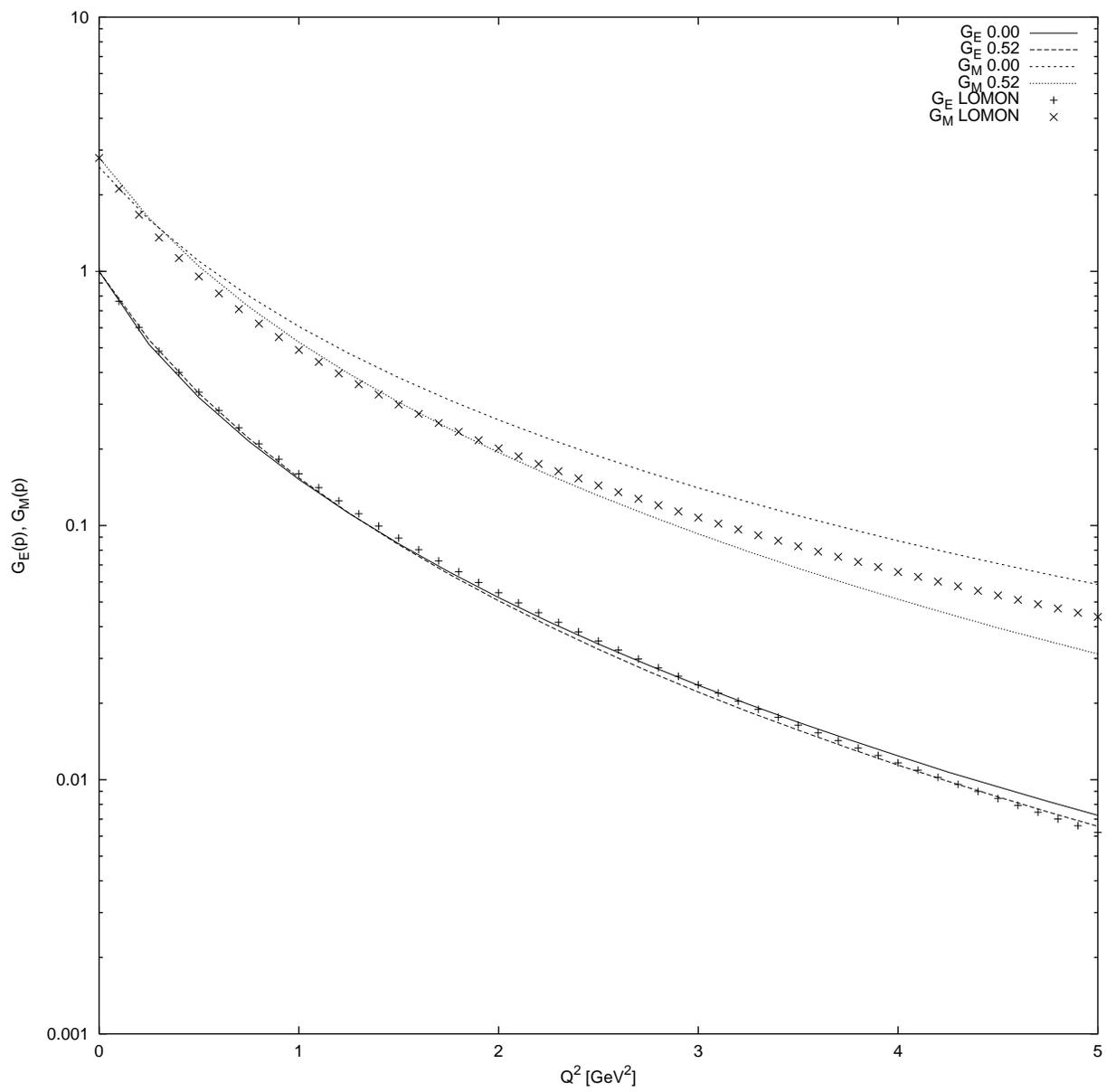
\end{document}